\documentclass[11pt]{article}
\usepackage{hyperref}
\pdfoutput=1
\begin{document}
\title{Spontaneous Formation of Columnar Vortices}
\author{Jared P. Whitehead$^1$, Beth A. Wingate$^1$ \\ \\
$^1$ Center for Nonlinear Studies (CNLS),\\ and Computational Physics and Methods (CCS2),\\ Los Alamos National Laboratory, Los Alamos, NM, USA}
\maketitle

\begin{abstract}
A fluid dynamics video of the rotating, weakly stratified Boussinesq equations is presented that illustrates the spontaneous formation of columnar vortices in the presence of stochastic, white noise forcing.
\end{abstract}

\section{Description of the video}
The rotating, stratified Boussinesq equations are a fundamental model of oceanic and atmospheric dynamics.  In the limit of rapid rotation (small Rossby $Ro$ number) the classical Taylor-Proudmann theorem indicates that the fluid motion is restricted to columnar structures \cite{WiEmHoTa2011}.  This video shows the formation of the vortex columns in one component of the horizontal velocity field.  The data is collected from a direct numerical simulation (DNS) from the LANL/Sandia DNS code (see \cite{WiEmHoTa2011} for details regarding the simulation and code) of the rotating, stratified Boussinesq equations with the rotation parameter Rossby number $Ro = 0.05$, and stratification parameter Froude number $Fr = 1$.  As explained in \cite{WiEmHoTa2011} the momentum is forced via a stochastic white-noise forcing centered at wave-number 3, and the initial conditions are zero for all dynamic quantities.  The computational grid is $256^3$ in order to resolve the smallest scale which for these simulations is dictated by the Rossby deformation radius (see \cite{WhWi2012} for a brief discussion).

The first part of the movie shows the effects of the stochastic forcing from the zero initial condition.  One can recognize large scale structures in the velocity field that are a result of the wave-number 3 dependence of the forcing.  Eventually the initial cutoff for the isosurfaces of the velocity field become obscured, so we renormalize these cutoffs.  In this intermediate time, we recognize the eventual organization of columnar structures in the flow field.  These columnar structures are paired, i.e. for each positive strong velocity column there is a corresponding negative velocity column directly above (in the sense of the video) it.  These structures correspond to a vortex column.  After the columnar vortices are fully formed, we renormalize the isosurfaces yet again so they are more visible.  The remainder of the video depicts the effective vortex dynamics as the vortices interact, dictating the flow.  This spontaneous evolution of vortex columns provides strong quantitative support of the classical Taylor-Proudmann theorem.

The computations were conducted on the Institutional Computing resources of the U. S. Department of Energy's Los Alamos National Laboratory.  We gratefully acknowledge the guidance and comments of J. Patchett, and the support of the U.S. Department of Energy's LANL/LDRD program.

\end{document}